\documentclass[twocolumn,aps,showpacs]{revtex4}
\usepackage{amssymb}
\usepackage{bm}
\usepackage{array}
\usepackage{graphicx}
\usepackage[indent]{caption2}
\usepackage{amsmath}
\usepackage{subfigure}
\usepackage[sort&compress]{natbib}

\begin{document}
  \title{Implementation of holonomic  quantum computation through
    engineering and manipulating environment}
\author{Zhang-qi Yin}
\author{Fu-li Li}
\email[Email: ]{flli@mail.xjtu.edu.cn}
\author{Peng Peng}
  \affiliation{Department of Applied Physics, Xi'an Jiaotong
  University, Xi'an 710049, China}

 \begin{abstract}
 We consider an atom-field coupled system, in which two pairs of four-level
 atoms are respectively driven by laser fields and trapped in two
 distant cavities that are connected by an optical fiber. First, we
 show that an effective squeezing reservoir can be engineered under
 appropriate conditions. Then, we show that a two-qubit geometric
 CPHASE gate between the atoms in the two cavities can be implemented
 through adiabatically manipulating the engineered reservoir along a
 closed loop. This scheme that combines engineering environment with
 decoherence-free space and geometric phase quantum computation
 together has the remarkable feature: a CPHASE gate with
 arbitrary phase shift is implemented by simply changing the
 strength and relative phase of the driving fields.
\end{abstract}
\pacs{03.67Lx, 03.65.Vf, 03.65.Yz, 42.50.Dv}
 \maketitle
\section{introduction}
Quantum computation, attracting much current interest since Shor's
algorithm \cite{Shor94} was proposed, depends on two key factors:
quantum entanglement and precision control of quantum systems.
Unfortunately, quantum systems  are inevitably coupled to their
environment so that entanglement is too fragile to be retained. This
makes the realization of quantum computation extremely difficult in
the real world. In order to overcome this difficulty, one proposed
the decoherence-free space concept \cite{DG97,ZR97}. It is found
that when qubits involved in quantum computation collectively
interact with a same environment there exists a ``protected''
subspace in the entire Hilbert space, in which the qubits are immune
to the decoherence effects induced by the environment. This subspace
is called decoherence-free space (DFS). To perform quantum
computation in a DFS, one has to design the specific Hamiltonian
containing controlling parameters, which eigenspace is spanned by
DFS states and the state-unitary manipulation related to quantum
computation goal is implemented by changing the controlling
parameters \cite{LCW98}.

As well known, instantaneous eigenstates of a quantum system with
the time-dependent Hamiltonian may acquire a geometric phase when
the time-dependent parameters adiabatically undergo a closed loop in
the parameter space \cite{Berry84}. The phase depends only on the
swept solid angle by the parameter vector in the parameter space.
This feature can be utilized to implement geometric quantum
computation (GQC) which is resilient to stochastic control errors
\cite{ZR99,DCZ01,ZZ05}. On combining the DFS approach with the GQC
scheme, one may build quantum gates which may be immune to both the
environment-induced decoherence effects and the control-led errors
\cite{WZL05}. In the scheme, quantum logical bits are represented by
degenerate eigenstates of the parameterized Hamiltonian. These
states have the features: they belong to DFS, and unitarily evolve
in time and acquire a geometric phase when the controlling
parameters adiabatically vary and undergo a closed loop.

In the recent paper \cite{CP+06}, Carollo and coworkers showed that
a cascade three-level atom interacting with a broadband squeezed
vacuum bosonic bath can be prepared in a state which is decoupled to
the environment. This state depends on the reservoir parameters such
as squeezing degree and phase angle. As the squeezing parameters
smoothly vary, the atomic state can unitarily evolve in time and
always be in the manifold of the DFS. Moreover, after a cyclic
evolution of the squeezing parameters, the state acquires a
geometric phase. This investigation has been generalized to cases
where both quantum systems and manipulated reservoir under
consideration are not restricted to cascade three-level atoms and
squeezed vacuum \cite{CSV06}. These results strongly inspire us that
instead of engineering Hamiltonian one may implement the
decoherence-free GQC by engineering and manipulating reservoir.

In this paper, we  propose a scheme in which the quantum-reservoir
engineering \cite{cirac92,MK+00,CP03} is combined with DFS and Berry
phase together to realize a two-qubit CPHASE gate \cite{Llo95}. We
show that atomic states can unitarily evolve in time in a DFS if the
change rate of reservoir parameters is much smaller than the
characteristic relaxation time of an atom-reservoir coupled system.
Moreover, we find that as the reservoir parameters adiabatically
change in time along an appropriate closed loop, the atomic state in
the DFS acquires a Berry phase and a CPHASE gate with arbitrary
phase shift can be realized. To our knowledge, it is the first
proposal for the realization of quantum gates by engineering and
steering the environment.

This paper is organized as follows. In Sec. II, we introduce a
cavity-atom coupling model in which two pairs of four-level atoms
are respectively trapped in two distant cavities that are connected
by an optical fiber. In the model, each of pairs of the atoms are
simultaneously driven by laser fields and coupled to the local
cavity modes through the double Raman transition configuration.
Under large detuning and bad cavity limits, we investigate to
engineer an effective broadband squeezing reservoir for the atoms.
In Sec. III, we analyze how to realize controlling gates between the
atoms trapped in the two cavities by steering the squeezing
reservoir. Section IV contains conclusions of our investigations.

\begin{figure}[htbp]
  \centering
  \includegraphics[width=6cm]{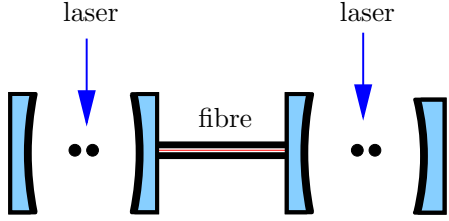}
  \caption{Atom-field coupling scheme.}
 \label{fig:scheme1}
\end{figure}
\begin{figure}[htbp]
  \centering
  \includegraphics[width=6cm]{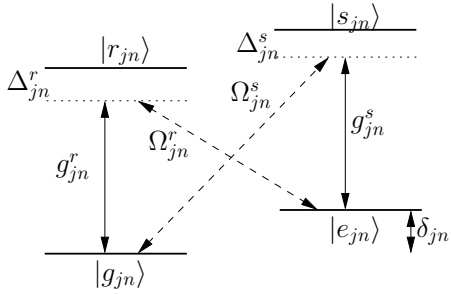}
  \caption{Atomic level
    configuration for atom $j$ in cavity $n$.}
 \label{fig:scheme2}
\end{figure}

\section{engineering a squeezing environment and generating a decoherence-free subspace}
Our scheme is shown in Fig.\ref{fig:scheme1}. A pair of four-level
atoms are trapped in each of two distant cavities, respectively,
which are connected through an optical fiber. In the short fiber
limit \cite{Pel97,SM+06,YL07}, only one fiber mode $b$ is excited
and coupled to cavity modes $a_1$ and $a_2$ with strength $\nu$
\cite{EK+99}. We assume that the cavity modes and the fiber mode
have the same frequency $\omega$. The level scheme of atoms is shown
in Fig.\ref{fig:scheme2}. Atom j in cavity $n$ is labeled by the
index $jn$ with $j,n=1,2$. The distance between the atoms in the
same cavity is assumed to be large enough that there is no direct
interaction between the atoms. The levels %$|a_{jn}\rangle$,
$|g_{jn}\rangle$ and $|e_{jn}\rangle$ of atom $j$ in cavity $n$,
with $j,n=1,2$ are stable with a long life time. The energy of the
level $|g_{jn}\rangle$ is taken to be zero as the energy reference
point. The lower lying level $|e_{jn}\rangle$, and upper levels
$|r_{jn}\rangle$ and $|s_{jn}\rangle$ have the energy $\delta_{jn}$,
and $\omega_{jn}^r$ and $\omega_{jn}^s$, respectively, in the unit
with $\hbar=1$. Transitions $|g_{jn}\rangle \leftrightarrow
|s_{jn}\rangle$ and $|e_{jn}\rangle \leftrightarrow |r_{jn}\rangle$
are driven by laser fields of frequencies $\omega^{L_s}_{jn}$ and
$\omega^{L_r}_{jn}$ with Rabi frequencies $\Omega^s_{jn}$ and
$\Omega^r_{jn}$ and relative phase $\varphi$, respectively.
Transitions $|g_{jn} \rangle \leftrightarrow |r_{jn}\rangle$ and
$|e_{jn}\rangle \leftrightarrow |s_{jn}\rangle$ are coupled to the
cavity mode $a_n$ with the strengths $g^r_{jn}$ and $g^s_{jn}$,
respectively. Here, we set $\Delta^r_{jn} = \omega^r_{jn} - \omega =
\omega^r_{jn} - \omega^{L_r}_{jn} - \delta_{jn}$, and
$\Delta^s_{jn}= \omega^s_{jn} -\omega - \delta_{jn} =\omega^{s}_{jn}
-  \omega^{L_s}_{jn}$.

Under the Markovian approximation, the master equation of the
density matrix for the whole system under consideration can be
written as \cite{CP03}
\begin{equation}
  \label{eq:M2master}
  \dot{\rho}_T = -i[H,\rho_T] + L_{cav_1}\rho_T+ L_{cav_2}\rho_T +
  L_{fiber}\rho_T,
\end{equation}
where $H= H_{0}+ H_{d} + H_{ac}+ H_{cf}$ with
\begin{equation}
  \label{eq:M2H}
  \begin{aligned}
   H_{0}=& \sum_{j,n=1}^2 \big( \omega_{jn}^{r} |r_{jn}\rangle \langle r_{jn}|
    + \omega^s_{jn} |s_{jn}\rangle \langle s_{jn} | + \delta_{jn} |e_{jn}\rangle
    \langle e_{jn} |) \\ &+
     \omega (\sum_{n=1}^2a^\dagger_n a_n+ b^\dagger b),\\
    H_{d} =& \sum_{j,n=1}^2 \big( \frac{\Omega^s_{jn}}{2} e^{-i\omega^{L_s}_{jn}
      t } |s_{jn}\rangle \langle g_{jn}| \\ &+ \frac{\Omega^r_{jn}}{2}
    e^{-i(\omega^{L_r}_{jn} t + \varphi)} |r_{jn} \rangle \langle e_{jn}| +
    \mathrm{H.c.} \big),  \\
    H_{ac}= & \sum_{j,n=1}^2 (g^r_{jn} |r_{jn}\rangle \langle g_{jn}|
    a_n + g^s_{jn} |s_{jn}\rangle \langle e_{jn}| a_n + \mathrm{H.c.}), \\
    H_{cf} = &\nu \Big[ b(a^{\dagger}_1 + a^{\dagger}_2 ) + \mathrm{H.c.}
    \Big].
  \end{aligned}
\end{equation}
Here, $H_0$ is the free energy of atoms and cavity fields, $H_d$ is
the interaction energy between the atoms and laser fields, $H_{ac}$
is the interaction energy between the atoms and the cavity fields,
and $H_{cf}$ describes the interaction between the cavity modes and
the fiber mode. The last three terms in \eqref{eq:M2master} describe
the relaxation processes of the cavity and fibre modes in the usual
vacuum reservoir, taking the forms
%\cite{cirac92,KC04}
\begin{equation}
  \label{eq:squres}
  \begin{aligned}
    L_{cav_n} \rho_T = &\kappa_n ( 2a_n \rho_T a^\dagger_n -
    a^\dagger_n a_n \rho_T - \rho_T a^\dagger_n a_n ), \\
    L_{fiber} \rho_T =& \kappa_f (2 b\rho_T b^\dagger - b^\dagger b \rho_T -
    \rho_T b^\dagger b),
  \end{aligned}
\end{equation}
where $\kappa_n$ is the leakage rate of photons from cavity $n$, and
$\kappa_f$ is the decay rate of the fiber mode.

Let's introduce collective basis: $|a\rangle_n = (|g_{1n}
\rangle|e_{2n}\rangle - |e_{1n}\rangle |g_{2n}\rangle )/\sqrt{2}$,
$|-1\rangle_n = |g_{1n}\rangle |g_{2n}\rangle$, $|0\rangle_n =
(|g_{1n}\rangle |e_{2n}\rangle + |e_{1n}\rangle |g_{2n}
\rangle)/\sqrt{2}$, $|1\rangle_n= |e_{1n} \rangle|e_{2n}\rangle $.
The states $|a\rangle_n$ and $|-1\rangle_n$ are taken as a qubit $n$
for quantum computation. In the large detuning limit, adiabatically
eliminating the excited states and setting $\frac{\Omega^r_{1n}
g^r_{1n}}{2\Delta^r_{1n}} = \frac{\Omega^r_{2n}
g^r_{2n}}{2\Delta^r_{2n}}= \beta^r_n$ and $\frac{\Omega^s_{1n}
g^s_{1n}}{2 \Delta^s_{1n}} = \frac{\Omega^s_{2n}
  g^s_{2n}}{2\Delta^s_{2n}} =\beta^s_n$, from
\eqref{eq:M2H}, we obtain the effective interaction Hamiltonian
\begin{equation}
  \label{eq:m2Heff}
  H_{eff} = \sum_{n} \sqrt{2}\big[  a_n (\beta^r_n e^{i\varphi} S^+_n
  + \beta^s_n S_n) + \mathrm{H.c.}  \big]+ H_{cf},
\end{equation}
where $S^+_n = |0\rangle_n {}_n\langle -1| + |1\rangle_n{}_n\langle
0|$. In the derivation of \eqref{eq:m2Heff}, we have assumed the
resonant condition $ \frac{{g^s_{jn}}^2}{ \Delta^s_{jn}} \langle
a^\dagger_n a_n\rangle + \frac{{\Omega^r_{jn}}^2}{4 \Delta^r_{jn}} =
\frac{{\Omega^s_{jn}}^2}{4\Delta^s_{jn}}+ \frac{{g^r_{jn}}^2}{
\Delta^r_{jn}} \langle a^\dagger_n a_n\rangle+\delta'_{jn}$. In
order to satisfy the condition with the flexible choice of
$\Omega^r_{jn}$, $\Omega^s_{jn}$, $\Delta^r_{jn}$ and
$\Delta^s_{jn}$, we have introduced additional ac-Stark shifts
$\delta'_{jn}$ to states $|g_{jn}\rangle$, which can be generated by
using a laser field to couple the level $|g_{jn}\rangle$ to an
ancillary level.

We now introduce three normal modes $c$ and $c_\pm$ with frequencies
$\omega$ and $\omega \pm \sqrt{2}\nu$ by use of the unitary
transformation $a_1 = \frac{1}{2} (c_+ + c_- + \sqrt{2} c)$,
    $a_2 = \frac{1}{2} (c_+ + c_- - \sqrt{2}c)$,
    $b =   \frac{1}{\sqrt{2}}( c_+ - c_-)$ \cite{SM+06,YL07}.
In the limit $\nu \gg |\beta^r_j|,|\beta^s_j|$, neglecting the far
off-resonant modes $c_\pm$ and setting $\beta^p_1= - \beta^p_2 =
\beta^p$ with $p=r,s$, we can approximately write the effective
Hamiltonian \eqref{eq:m2Heff} as
\begin{equation}
  \label{eq:Heff1}
  H_{eff} = (\beta^r e^{i\varphi} S^+ + \beta^s S) c + \mathrm{H.c.},
\end{equation}
where $S^+=S^+_1+S^+_2$.

Since the modes $c_\pm$ are nearly not excited and decoupled with
the resonant mode $c$, the fiber mode $b$ is mostly in the vacuum
state, therefore, $L_{fiber}\rho_T$ can be neglected, and
$L_{cav_1}\rho_T + L_{cav_2} \rho_T$ can be approximated as
\begin{equation}
  \label{eq:M2squc}
   L_{cav} \rho_T = \kappa ( 2c \rho_T c^\dagger -
    c^\dagger c \rho_T - \rho_T c^\dagger c ),
\end{equation}
where $\kappa=(\kappa_1+\kappa_2)/2$.

In the bad cavity limit, $\kappa \gg \beta$, adiabatically
eliminating the mode $c$ \cite{cirac92,CP03}, from Eq.
\eqref{eq:M2master} with the replacement of the Hamiltonian
\eqref{eq:M2H} and the relaxation terms \eqref{eq:squres} by the
effective Hamiltonian \eqref{eq:Heff1} and the relaxation term
\eqref{eq:M2squc}, respectively, we can obtain the master equation
for the density matrix of the atoms
\begin{equation}
  \label{eq:M2rhoa1}
  \dot{\rho} = - \frac{\Gamma}{2}( R^+ R \rho + \rho R^+ R -
  2R \rho R^+),
\end{equation}
where $\rho= \mathrm{Tr}_f (\rho_T)$, $R= S \cosh r + e^{i\varphi}
S^\dagger \sinh r$, $r= \cosh^{-1} (\beta^r/\sqrt{{\beta^r}^2 -
  {\beta^s}^2})$ and $\Gamma = 2({\beta^r}^2 - {\beta^s}^2)/\kappa$.
Eq. \eqref{eq:M2rhoa1} describes the collective interaction of two
cascade three-level atoms with the effective squeezed vacuum
reservoir \cite{CP+06}.  The parameters $\beta^r$, $\beta^s$ and
$\varphi$ are easily changed and controlled at will by varying the
strength and phase of the driving lasers \cite{DCZ01}. We will show
that a geometric phase gate can be realized through changing these
parameters.

The DFS of the atomic system is spanned by the states which satisfy
the equation $R(r,\varphi) |\psi_{\mathrm{DF}}(r,\varphi)\rangle =0$
\cite{CP+06}. In terms of basis states
   $ |e_1\rangle = |a\rangle_1 |a\rangle_2, ~
    |e_2\rangle = |a\rangle_1 |-1\rangle_2,~
    |e_3\rangle = |-1\rangle_1 |a\rangle_2,~
    |e_4\rangle = |a\rangle_1 |0\rangle_2, ~
    |e_5\rangle = |0\rangle_1 |a\rangle_2, ~
    |e_6\rangle = |a\rangle_1 |1\rangle_2, ~
    |e_7\rangle = |1\rangle_1 |a\rangle_2,~
 |e_8\rangle = |1\rangle_1 |1\rangle_2, ~
  |e_9\rangle = \frac{1}{\sqrt{2}} (|1\rangle_1 |0\rangle_2 +
  |0\rangle_1 |1\rangle_2), ~
  |e_{10}\rangle = |-1\rangle_1 |-1\rangle_2, ~
  |e_{11}\rangle =  \frac{1}{\sqrt{2}} (|0\rangle_1 |-1\rangle_2 +
  |-1\rangle_1 |0\rangle_2), ~
  |e_{12}\rangle =  \frac{1}{\sqrt{6}} (|1\rangle_1 |-1\rangle_2 +
  |-1\rangle_1 |1\rangle_2 ) + \frac{2}{\sqrt{6}} |0\rangle_1
  |0\rangle_2)$, the DFS states can be written as
\begin{equation}
  \label{eq:DF}
 \begin{aligned}
     %|\psi_{\mathrm{DF}}(r,\varphi) \rangle_0 =&  |e_{13}\rangle,
     |\psi_{\mathrm{DF}}(r,\varphi) \rangle_1 =& |e_1\rangle, \\
  |\psi_{\mathrm{DF}}(r,\varphi) \rangle_j = &\frac{\cosh r}
  {\sqrt{\cosh 2r}} |e_j\rangle -  e^{i \varphi} \frac{\sinh
    r}{\sqrt{\cosh 2r}} |e_{j+4}\rangle, ~ j=2,3,\\
  |\psi_{\mathrm{DF}}(r,\varphi) \rangle_4 = &\frac{e^{2i\varphi} (\tanh r)^2
  |e_8\rangle - \sqrt{\frac{2}{3}} e^{i\varphi}\tanh r
  |e_{12}\rangle + |e_{10}\rangle}{ \sqrt{ (\tanh
    r)^4 + \frac{2}{3} (\tanh r)^2 + 1}}.
 \end{aligned}
\end{equation}

Let's introduce a unitary transformation $O(r,\varphi)$ by $|\phi_i \rangle
=\sum_{i=1}^{12} O_{ij}(r,\varphi) |e_j \rangle$, where $|\phi_i
\rangle= |\psi_{\mathrm{DF}}\rangle_i$ for $i=1,2,3,4$. For the
transformed density matrix $\bar{\rho} = O^\dagger \rho O$, we have
\begin{equation}
  \label{eq:master2}
  \frac{d\bar{\rho}}{dt} = i [G,\bar{\rho}] + O^\dagger
  \frac{d\rho}{dt} O,
\end{equation}
where $G(r,\varphi) = i O^\dagger\frac{dO}{dt} = i O^\dagger
[\dot{r} \frac{dO}{dr} +\dot{\varphi} \frac{dO}{d\varphi}]$. To
solve Eq. \eqref{eq:master2} in the DFS, let's define the
time-independent projector  $\Pi(0)=O^\dagger \Pi(r,\varphi) O =
\sum_{i=1}^4 O^\dagger |\phi_i\rangle \langle \phi_i|O =
\sum_{j=1}^3|e_j\rangle\langle e_j| + |e_{10}\rangle \langle
e_{10}|$ onto the DFS. From \eqref{eq:master2}, we obtain the
equation of motion for $\bar{\rho}_{\mathrm{DF}} = \Pi(0) \bar{\rho}
\Pi(0)$
\begin{equation}
  \label{eq:master3}
 \begin{aligned}
  \frac{d \bar{\rho}_{\mathrm{DF}}}{dt} = &i [ G_{\mathrm{DF}} ,
    \bar{\rho}_{\mathrm{DF}}] + i\Pi(0) G \Pi_\bot(0)
    \bar{\rho} \Pi(0)  \\ &  - i \Pi(0) \bar{\rho}
    \Pi_\bot (0) G \Pi(0)+  \Pi(0) O^\dagger \frac{d\rho}{dt} O \Pi(0),
  \end{aligned}
\end{equation}
where $\Pi_\bot(0) = \mathbf{1} - \Pi (0)$ and $G_{\mathrm{DF}} =
\Pi(0) G \Pi(0)$. In the limit of $\dot{r}, \dot{\varphi} \ll
\Gamma$, the last three terms in Eq. \eqref{eq:master3} can be
neglected \cite{CSV06}. In this way,  Eq. \eqref{eq:master3} is
reduced to
\begin{equation}
  \label{eq:rhoDF}
  \frac{d \bar{\rho}_{\mathrm{DF}}}{d t}  = i [ G_{\mathrm{DF}} ,
    \bar{\rho}_{\mathrm{DF}}].
\end{equation}
Therefore, in the frame dragged adiabatically by the reservoir, the
state of the atoms in the DFS unitarily evolves in time.
\section{realizing controlling phase gates
through manipulating the squeezing environment} In this section, we
investigate how to realize a CPHASE gate through manipulating the
engineered reservoir. Suppose that at the initial time the laser
field driving the transition $|g\rangle \leftrightarrow |s\rangle$
is switched off but the laser field  driving the transition
$|r\rangle \leftrightarrow |e\rangle$ is switched on and the atoms
are in the DFS state $|\Psi(0)\rangle_a = \frac{1}{2}(|a\rangle_1
|a\rangle_2 + |a\rangle_1 |-1\rangle_2 + |-1\rangle_1 |a\rangle_2 +
|-1\rangle_1 |-1\rangle_2) = \sum_{j=1}^4 |\psi_{ \mathrm{DF}}
(0,0)\rangle_j/2$. To generate a geometric phase for the atomic
state, we smoothly change the parameters of the engineered reservoir
along a closed loop, which is divided into the following three
steps: (1) From time $0$ to $T_1$, hold on $\varphi=0$, and
adiabatically increase the parameter $r$ from $0$ to $r_0$; (2) From
time $T_1$ to $T_2$, hold on $r=r_0$, and adiabatically change the
phase $\varphi$ from $0$ to $\varphi_0$; (3) From time $T_2$ to
$T_3$, hold on $\varphi= \varphi_0$, and
  adiabatically decrease $r$ from $r_0$ to $0$.
When the cyclic evolution ends,  the atomic state becomes
\begin{equation}
  \label{eq:atomstate}
 \begin{aligned}
  |\Psi(T_3)\rangle_a = \frac{1}{2}
  (|e_1\rangle + e^{i\chi_1}|e_2\rangle+ e^{i\chi_1} |e_3\rangle +
  e^{i\chi_{12}} |e_{10}\rangle),
  \end{aligned}
\end{equation}
where geometric phases $\chi_1= - \nu_1 \varphi_0$, $\chi_{12}=
-\nu_{12} \varphi_0$ with $\nu_1= \frac{\sinh^2 r_0 }{\sinh^2 r_0 +
  \cosh^2 r_0}$, $\nu_{12} = \frac{2 \tanh^4 r_0+ \frac{2}{3} \tanh^2
  r_0}{\tanh^4 r_0 + \frac{2}{3} \tanh^2 r_0 + 1}$.
By performing local transformations $U_1 = e^{-i\chi_1}|-1\rangle_1
{}_1 \langle -1|$ and $U_2= e^{-i\chi_1}|-1\rangle_2 {}_2 \langle
-1|$, the state \eqref{eq:atomstate} can be written as
$|\Psi'(T_3)\rangle_a = U_1U_2 |\Psi(T_3)\rangle_a = \frac{1}{2}
(|a\rangle_1|a\rangle_2 + |a\rangle_1 |-1\rangle_2 + |-1\rangle_1
|a\rangle_2 + e^{i\Delta} |-1\rangle_1 |-1\rangle_2)$, where
$\Delta= \chi_{12} - 2\chi_{1}= (2\nu_1 - \nu_{12}) \varphi_0$.
Thus, the CPHASE gate with the phase shift $\Delta$ is realized. If
both the atoms in cavity  $1$ and the atoms in cavity $2$ ``see''
different environments, $|\chi_{12}|$ must be equal to $|2\chi_1|$
and $\Delta=0$. Therefore, the phase shift $\Delta$ results from the
collective coupling of the atoms in both cavities with the same
engineered environment. If $r_0= \mathrm{atanh} ( \sqrt{
  \sqrt{4/3} -1})\simeq 0.4157$, $|\nu_{12}| = |\nu_{1}|$.
Under this condition with $\varphi_0=\pi/\nu_1$, the state of the
atoms at the time $T_3$ is $ |\Psi''(T_3)\rangle_a = -\frac{1}{2}
(-|a\rangle_1|a\rangle_2 + |a\rangle_1 |-1\rangle_2 + |-1\rangle_1
|a\rangle_2 + |-1\rangle_1 |-1\rangle_2) $. In this case, the
Controlled-Z gate between the two qubits is realized without local
transformations.

\begin{figure}[htbp]
  \centering
\includegraphics[width=8cm]{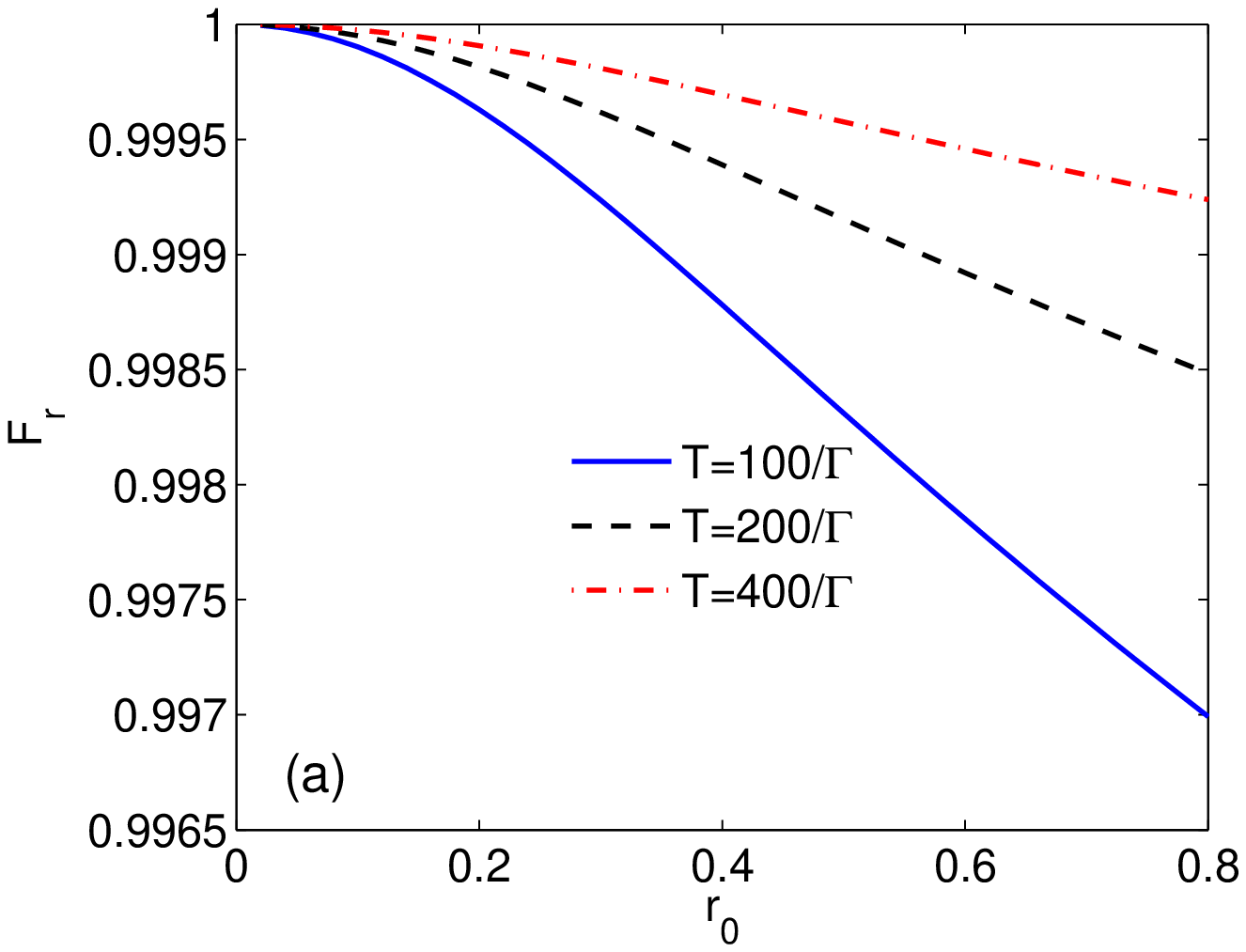}
%\onelinecaptionsfalse
\caption{Fidelity $F_r$ of the atomic state.}
\label{fig:BerryPhase1}
\end{figure}
\begin{figure}[htbp]
  \centering
\includegraphics[width=8cm]{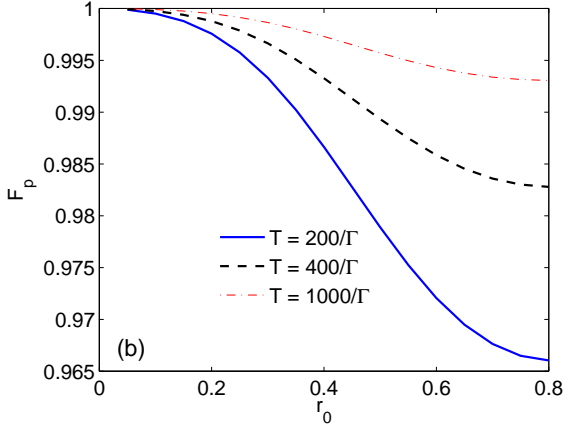}
%\onelinecaptionsfalse
\caption{Fidelity $F_p$ of the atomic state. }
\label{fig:BerryPhase2}
\end{figure}

The above results depend on the adiabatical approximation. To check
the adiabatical condition, we numerically simulate the following two
examples. In the first example, we suppose that at the initial time
the atoms are in the state $|\Psi_1\rangle_a = (|a\rangle_1
|a\rangle_2 + |\psi_{\mathrm{DF}} (0,0)\rangle_{2})/\sqrt{2}$ and
the laser field driving the transition $|e\rangle \leftrightarrow
|r\rangle$ are turned on. Then, by slowly switching the laser field
driving the transition $|g\rangle \leftrightarrow |s\rangle$, we
increase the parameter $r$ from $0$ to $r_0$ according to the linear
function $r(t) = r_0 t/T$. In the adiabatical limit $(T \gg
\Gamma^{-1})$, the atomic state becomes $|\Psi'_1 \rangle_a =
(|a\rangle_1 |a\rangle_2 + |\psi_{\mathrm{DF}} (r_0,0) \rangle_{2})
/\sqrt{2}$ at the time $T$. On the other hand, in the Hilbert space
spanned by the basis states $\{ |e_i\rangle \}$ for
$i=1,2,\cdots,12$ , we can numerically solve Eq. \eqref{eq:M2rhoa1}
and obtain the density matrix $\rho_1(T)$ of the atoms. Let's define
$F_r = {}_a\langle \Psi'_1|\rho_1(T) | \Psi'_1 \rangle_a $ as the
fidelity for this process. As shown in Fig. \ref{fig:BerryPhase1},
if $T> 100/\Gamma$, $F_r$ is always bigger than $0.997$ if $r \in (
0,0.8 )$, corresponding to the almost perfect evolution.

In the second example, we suppose that the atoms are initially in
the state $|\Psi_2\rangle_a = (|a\rangle_1 |a\rangle_2 +
|\psi_{\mathrm{DF}} (r,0)\rangle_{4})/\sqrt{2}$ and all the driving
fields are turned on to hold the parameters $r=r_0$ and $\varphi=0$.
By adiabatically changing the phase $\varphi$ from $0$ to $2\pi$ at
the rate $\dot{\varphi}= 2\pi/T$, the atomic state at the time $T$
becomes $|\Psi'_2 \rangle_a = (|a\rangle_1 |a\rangle_2 +
e^{i\chi_{12}} |\psi_{\mathrm{DF}} (r,2\pi) \rangle_{4}) /\sqrt{2}$.
Let's define the fidelity for this example as $F_p = {}_a\langle
\Psi'_2|\rho(T) | \Psi' \rangle_a $, where $\rho(T)$ is the
numerical solution of Eq. \eqref{eq:M2rhoa1}. As shown in Fig.
\ref{fig:BerryPhase2}, $F_p$ increases as $T$ increases but
decreases as the parameter $r_0$ increases. If $T>1000/\Gamma$,
$F_p$ is larger than $0.992$ for $0<r_0<0.8$. From these two
examples, we find that to fulfill the adiabatical condition the time
used in the step $2$ should be much longer than in the steps $1$ and
$3$.

A controlled-Z gate has been numerically simulated by directly
solving Eq. (7) with $r_0 = 0.5$, and $\varphi_0= \pi
/|2\nu_1-\nu_{12}|$. In the simulation, we set $\dot{r}=r_0/T_1$ in
the steps $1$ and $3$, and $\dot{\varphi}= \varphi_0/(T_2-T_1)$ in
the step $2$ with $T_1=0.05T_3$ and $T_2-T_1= 0.90T_3$. If
$T_3>1100/ \Gamma$, we find that the fidelity $F = {}_a \langle
\Psi(T_3) | \rho(T_3)| \Psi(T_3)\rangle_a$ is larger than $0.95$.
For an almost perfect controlled-Z gate with $F>0.99$, we find that
$T_3$ must be longer than $6000/ \Gamma$.

Now let's briefly discuss the effects of the atomic spontaneous
emission, the fiber mode decay and cavity photon leakage. For
simplicity but without the loss of generality, we suppose that
atomic spontaneous emission rates of the excited levels are equal to
$\gamma$. In the large detunig limit, the characteristic spontaneous
emission rate of the atoms is $\gamma_{eff} = \gamma( \Omega^2/2
\Delta^2)$ \cite{Pel97,CP03} and the effective decay rate of the
fiber mode is $\kappa_{eff}= \kappa_f \Omega^2 g^2/(4\Delta^2
\nu^2)$. If $\kappa_f \leq \gamma$ and $g^2 \ll \nu^2$,
$\kappa_{eff}$ can be much smaller than $\gamma_{eff}$. Under this
condition, the present scheme is feasible if $\Gamma \gg
\gamma_{eff}$. In the current cavity quantum dynamic (CQED)
experiment, the parameters $(g,\kappa, \gamma) = (2000, 10, 10)$ MHz
could be available \cite{SKV+05}. If setting $\Omega/ (2\Delta)=
\frac{1}{\sqrt{2}}\times10^{-3}$, we have $\Gamma \simeq 4
\times10^{4} \gamma_{eff}$. The condition is held. In the present
scheme, the large cavity decay rate is required to ensure that the
cavity modes are in a broadband squeezed vacuum reservoir and then
the atoms always "see" the broadband squeezed vacuum reservoir
during the dynamic evolution. For an arbitrary small but nonzero
value of the squeezing degree of the reservoir, a CPHASE gate with
arbitrary high fidelity can always be realized in the represent
scheme. The cavity decay does not directly affect the fidelity of
the realized CPHASE gates. However, the larger the decay rate is,
the longer the operation time of the CPHASE gates is. Thus, we have
the condition $\kappa>>\beta, \gamma_{eff}$ for realizing the
reliable CPHASE gates. Based on the parameters quoted above, this
condition can be well satisfied. With the parameters of the current
CQED experiment, we find that the operation time of the controlled-Z
gate, with fidelity larger than $0.95$, is about $2.8$ ms. It is
much shorter than both $1/\gamma_{eff}$ and the single-atom trapping
time in cavity \cite{NM+05}.  On the other hand, the present scheme
needs a strong coupling between the cavity and the fiber. This could
be realized at the current experiment \cite{SK+03}. Therefore, the
requirement for the realization of the present scheme can be
satisfied with the current technology.
\section{conclusions}
We propose a cavity-atom coupled scheme for the realization of
quantum controlling gates, in which each of two pairs of four-level
atoms in two distant cavities connected by a short optical fibre are
simultaneously driven by laser fields and coupled to the local
cavity modes through the double Raman transition configuration. We
show that an effective squeezing reservoir coupled to the multilevel
atoms can be engineered under appropriate driving condition and bad
cavity limit. We find that in the scheme a CPHASE gate with
arbitrary phase shift can be implemented through adiabatically
changing the strength and phase of driving fields along a closed
loop. It is also noticed that the larger the effective coupling
strength between the environment and the atoms is, the more reliable
the realized CPHASE gate is.

\acknowledgments We thank Yun-feng Xiao and Wen-ping He for valuable
discussions and suggestions. This work was supported by the Natural
Science Foundation of China (Grant Nos. 10674106, 60778021 and
05-06-01).

\bibliographystyle{apsrev}
%\bibliography{bibtex}

\end{document}